\documentclass[aps,twocolumn]{revtex4}
\usepackage{graphicx}

\begin{document}

\title{Simulation study of the interaction between large-amplitude HF radio
waves and the ionosphere}

\author{B. Eliasson}
\affiliation{Department of Physics, Ume{\aa} University, SE-901 87
Ume{\aa}, Sweden} \affiliation{Theoretische Physik IV,
Ruhr--Universit\"at Bochum, D-44780 Bochum, Germany}

\author{B. Thid\'e}
\affiliation{ Swedish Institute of Space Physics, P. O. Box 537,
SE-751 21 Uppsala, Sweden } \affiliation{LOIS Space Centre,
V\"axj\"o University, SE-351 95 V\"axj\"o, Sweden}

\begin{abstract}
The time evolution of a large-amplitude electromagnetic (EM) wave
injected vertically into the overhead ionosphere is studied
numerically. The EM wave has a carrier frequency of 5 MHz and is
modulated as a Gaussian pulse with a width of approximately 0.1
milliseconds and a vacuum amplitude of 1.5 V/m at 50 km. This is a
fair representation of a modulated radio wave transmitted from a
typical high-power HF broadcast station on the ground. The pulse is
propagated through the neutral atmosphere to the critical points of
the ionosphere, where the L-O and R-X modes are reflected, and back
to the neutral atmosphere. We observe mode conversion of the L-O
mode to electrostatic waves, as well as harmonic generation at the
turning points of both the R-X and L-O modes, where their amplitudes
rise to several times the original ones. The study has relevance for
ionospheric interaction experiments in combination with ground-based
and satellite or rocket observations.
\end{abstract}

\maketitle
%% ------------------------------------------------------------------------ %%
%
%  BEGIN ARTICLE
%
%% ------------------------------------------------------------------------ %%

% The body of the article must start with a \begin{article} command,
% and an \end{article} command must be placed at the end of the file,
% before \end{document}.
%
% If using draft mode \end{article} must follow the references section.

%\begin{article}

%% ------------------------------------------------------------------------ %%
%
%  TEXT
%
%% ------------------------------------------------------------------------ %%

\section{Introduction}
Pulsed high-frequency (HF) electromagnetic (EM) waves from
transmitters on the ground are regularly used for sounding the
density profile and drift velocity of the overehead ionosphere
[\textit{Hunsucker}, 1991; \textit{Reinisch et al.}, 1995,
\textit{Reinisch}, 1996]. In 1971, it was shown theoretically by
\textit{Perkins and Kaw} [1971] that if the injected HF radio beams
are strong enough, weak-turbulence parametric instabilities in the
ionospheric plasma of the type predicted by \textit{Silin} [1965]
and \textit{DuBois and Goldman} [1965] would be excited. Ionospheric
modification experiments by a high-power HF radio wave at
Platteville in Colorado [\textit{Utlaut}, 1970], using ionosonde
recordings and photometric measurements of artificial airglow,
demonstrated the heating of electrons, the deformation in the traces
on ionosonde records, the excitation of spread $F$, etc., after the
HF transmitter was turned on. The triggering of weak-turbulence
parametric instabilities in the ionosphere was first observed in
1970 in experiments on the interaction between powerful HF radio
beams and the ionospheric plasma, conducted at Arecibo, Puerto Rico,
using a scatter radar diagnostic technique [\textit{Wong and
Taylor}, 1971; \textit{Carlson et al.}, 1972].
%Theoretical and experimental work of
%stimulated Brillouin scattering involved also the EISCAT and
%Jicamarca facilities [\textit{Larsson et al.}, 1976; \textit{Fejer},
%1977; \textit{Dysthe et al.}, 1977; \textit{Fejer et al.}, 1978].
A decade later it was found experimentally in Troms{\o} that, under
similar experimental conditions as in Arecibo, strong, systematic,
structured, wide-band secondary HF radiation escapes from the
interaction region [\textit{Thid\'e et al.}, 1982]. This and other
observations demonstrated that complex interactions, including weak
and strong EM turbulence, [\textit{Leyser}, 2001; \textit{Thid\'e et
al.}, 2005] and harmonic generation [\textit{Derblom et al.}, 1989;
\textit{Blagoveshchenskaya et al.}, 1998] are excited in these
experiments.
%Some recent results are presented by Thid\'e et al.,
%[2005].
% By varying the injected HF beam in terms of
%frequency, intensity, beam direction, sense of polarization, and
%duty cycle and analyzing the secondary radiation, it has been
%possible to study systematically, at different times and different
%sites, the ionospheric plasma turbulence and wave conversion
%processes.
%The heating
%of electrons also leads to optic emissions, or artificial airglow,
%from the upper-hybrid layer when the frequency of the transmitter is
%close to that of one of the electron cyclotron harmonics
%[\textit{Djuth et al.}, 2005; \textit{Gustavsson et al.}, 2006].

Numerical simulations have become an important tool to understand
the complex behavior of plasma turbulence. Examples include
analytical and numerical studies of Langmuir turbulence
[\textit{Robinson}, 1997], and of upper-hybrid/lower-hybrid
turbulence in magnetized plasmas [\textit{Goodman et al.}, 1994;
\textit{Xi}, 2004].
% and indicate that possible mechanisms for the
%broad upshifted maximum involve a four-wave decay and non-Maxwellian
%electrons [\textit{Hussein et al.}, 1998; \textit{Xi}, 2004].
In this Letter, we present a full-scale simulation study of the
propagation of an HF EM wave into the ionosphere, with ionospheric
parameters typical for the high-latitude EISCAT Heating facility in
Troms{\o}, Norway. To our knowledge, this is the first simulation
involving realistic scale sizes of the ionosphere and the wavelength
of the EM waves. Our results suggest that such simulations, which
are possible with today's computers, will become a powerful tool to
study HF-induced ionospheric turbulence and secondary radiation on a
quantitative level for direct comparison with experimental data.
\section{Mathematical Model and Numerical Setup}
We use the MKS system (SI units) in the mathematical expressions
throughout the manuscript, unless otherwise stated. We assume a
vertically stratified ion number density profile $n_{i0}(z)$ with a
constant geomagnetic field ${\bf B}_0$ directed obliquely to the
density gradient. The EM wave is injected vertically into the
ionosphere, with spatial variations only in the $z$ direction. Our
simple one-dimensional model neglects the EM field $1/r$ falloff
($r$ is the distance from the transmitter), the Fresnel pattern
created obliquely to the $z$ direction by the incident and reflected
wave, and the the influence on the radio wave propagation due to
field aligned irregularities in the ionosphere. For the EM wave, the
Maxwell equations give
\begin{equation}
\frac{\partial {\bf B}_1}{\partial t}=-\widehat{\bf z}\times
\frac{\partial{\bf E}}{\partial z},
\end{equation}
\begin{equation}
\frac{\partial {\bf E}}{\partial t}=c^2\widehat{\bf z}\times
\frac{\partial {\bf B}_1}{\partial z}+\frac{en_{\rm e}{\bf v}_{\rm
e}}{\varepsilon_0},
\end{equation}
where the electron fluid velocity is obtained from the momentum
equation
\begin{equation}
\frac{\partial {\bf v}_{\rm e}}{\partial t}=-v_{{\rm
e}z}\frac{\partial {\bf v}_{\rm e}}{\partial z}-\frac{e}{m_{\bf
e}}[{\bf E}+{\bf v}_{\rm e}\times ({\bf B}_0+{\bf B}_1)]
\end{equation}
and the electron density is obtained from the Poisson equation
$n_{\rm e}=n_{\rm i0}(z)-({\varepsilon_0}/{e}){\partial
E_z}/{\partial z}$. Here, $\widehat{\bf z}$ is the unit vector in
the $z$ direction, $c$ is the speed of light in vacuum, $e$ is the
magnitude of the electron charge, $\varepsilon_0$ is the vacuum
permittivity, and $m_{\rm e}$ is the electron mass.

\begin{figure}[htb]
\includegraphics[width=0.48\textwidth]{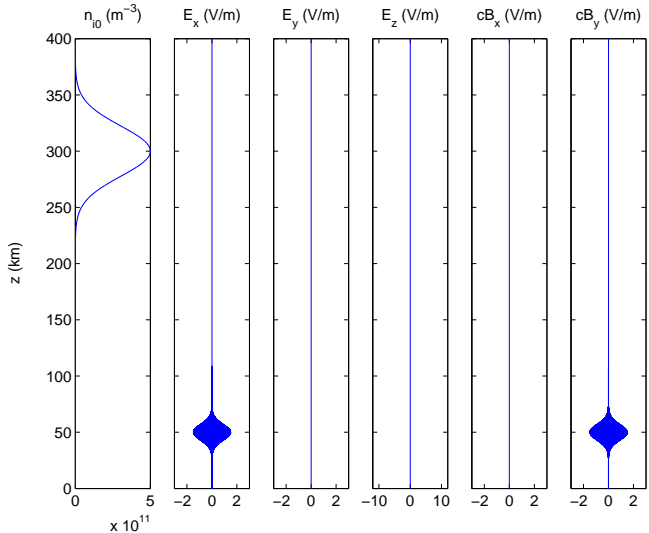}
\caption{The ion density profile, and the electric and magnetic
field components at time $t=0$ ms.} \label{Fig1}
\end{figure}

The number density profile of the immobile ions,
$n_{i0}(z)=0.5\times 10^{12}\exp[-(z-300)^2/10^3]$ ($z$ in
kilometers) is shown in the leftmost panel of Fig.~\ref{Fig1}.
Instead of modeling a transmitting antenna via a time-dependent
boundary condition at $z=0$ km, we assume that the EM pulse has
reached the altitude $z=50$ km when we start our simulation, and we
give the pulse as an initial condition at time $t=0$ s. In the
initial condition, we use a linearly polarized EM pulse where the
carrier wave has the wavelength $\lambda=60\,{\rm m}$ (wavenumber
$k=0.1047\,{\rm m}^{-1}$) corresponding to a carrier frequency of
$f_0=5\,{\rm MHz}$ ($\omega_0=31\times10^6\,{\rm s}^{-1}$). The EM
pulse is amplitude modulated in the form of a Gaussian pulse with a
maximum amplitude of $1.5$ V/m, with the $x$-component of the
electric field set to $E_x=1.5\exp[-(z-50)^2/10^2]\sin(0.1047\times
10^{3} z)$ ($z$ in kilometers) and the $y$ component of the magnetic
field set to $B_y=E_x/c$ at $t=0$.  The other electric and magnetic
field components are set to zero; see Fig.~\ref{Fig1}. The spatial
width of the pulse is approximately 30 km, corresponding to a
temporal width of 0.1 milliseconds as the pulse propagates with the
speed of light in the neutral atmosphere. It follows from Eq.~(1)
that $B_z$ is time-independent; hence we do not show $B_z$ in the
figures. The geomagnetic field is set to $B_0=4.8\times10^{-5}$
Tesla, corresponding to an electron cyclotron frequency of 1.4 MHz,
directed downward and tilted in the $xz$-plane with an angle of $13$
degrees ($0.227$ rad) to the $z$-axis, i.e., ${\bf
B}_0=(B_{0x},B_{0y},B_{0z})=(\sin 0.227,0,-\cos 0.227)B_0$. In our
numerical simulation, we use $10^5$ spatial grid points to resolve
the plasma for $0\leq z \leq 400$ km. The spatial derivatives are
approximated with centered second-order difference approximations,
and the time-stepping is performed with a leap-frog scheme with a
time step of $\Delta t=8\times 10^{-9}$~s.
\section{Numerical Results}
\begin{figure}[htb]
\includegraphics[width=0.48\textwidth]{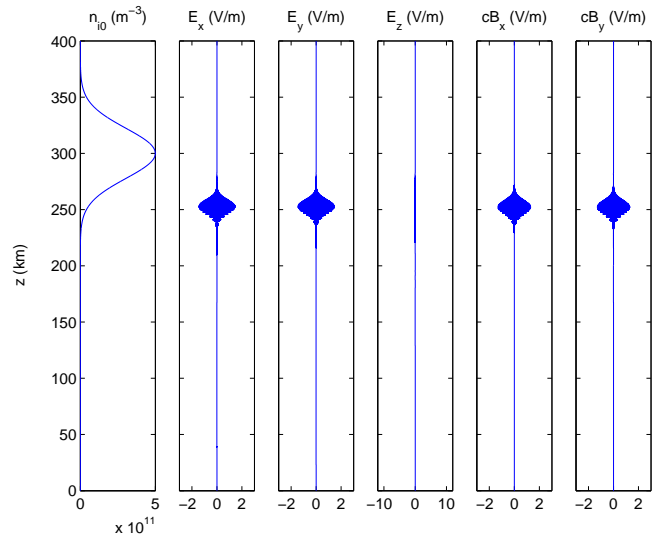}
\caption{The ion density profile, and the electric and magnetic
field components at time $t=0.720$ ms. The splitting of the wave is
due to Faraday rotation.} \label{Fig2}
\end{figure}

\begin{figure}[htb]
\includegraphics[width=0.48\textwidth]{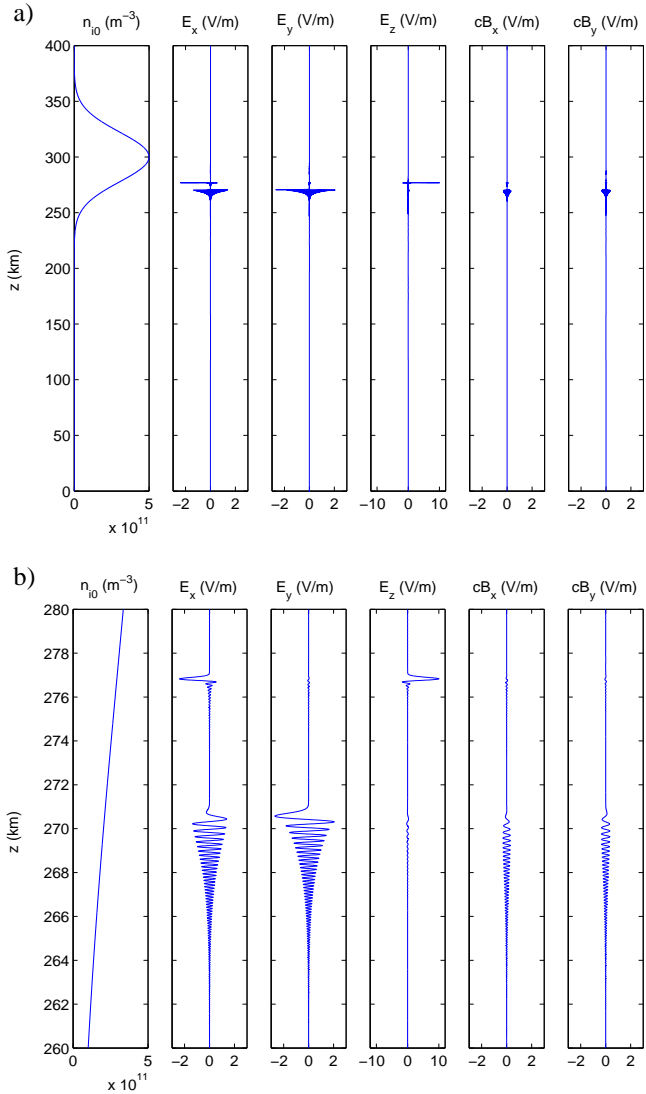}
\caption{a) The ion density profile, and the electric and magnetic
field components at time $t=0.886$ ms. b) A closeup of the region of
the turning points of the R-X and L-O modes. We see that the
wave-energy of the L-O mode is concentrated into one single
half-wave envelop at $z\approx 277$ km, while the turning point of
the less localized R-X mode is at $z\approx 270.5$ km.}
\label{Fig3-4}
\end{figure}

\begin{figure}[htb]
\includegraphics[width=0.48\textwidth]{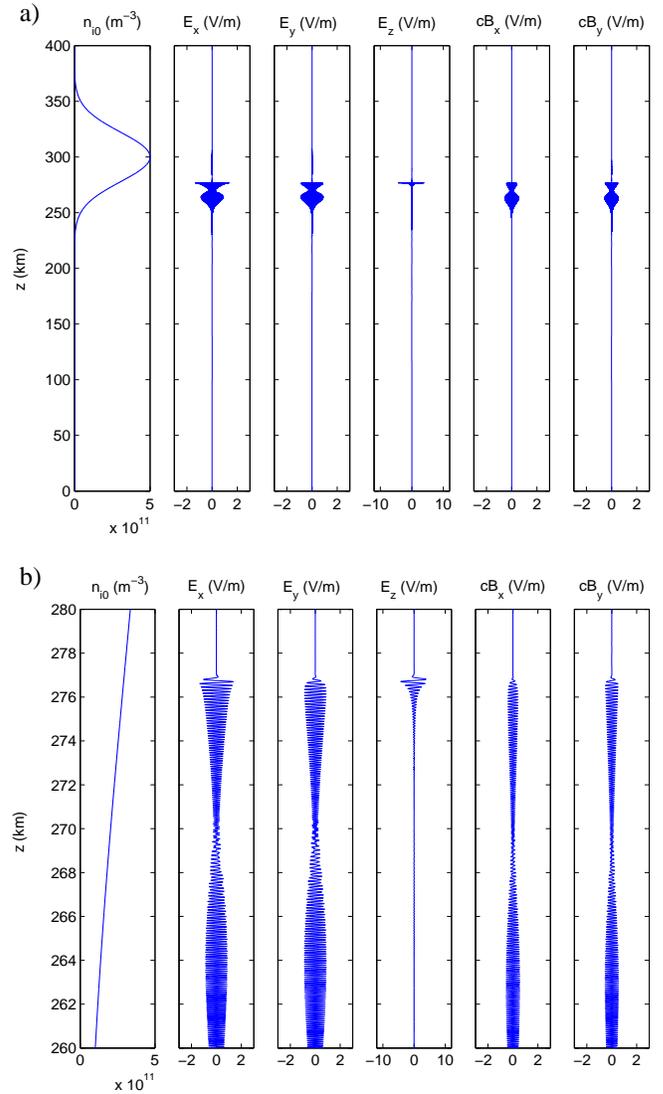}
\caption{a) The ion density profile, and the electric and magnetic
field components at time $t=0.948$ ms. b) A closeup of the region of
the turning points of the R-X and L-O modes. Here, the L-O mode
oscillations at $z\approx 277$ km are radiating EM waves with
perpendicular (to the $z$ axis) electric field components.}
\label{Fig5-6}
\end{figure}

\begin{figure}[htb]
\includegraphics[width=0.48\textwidth]{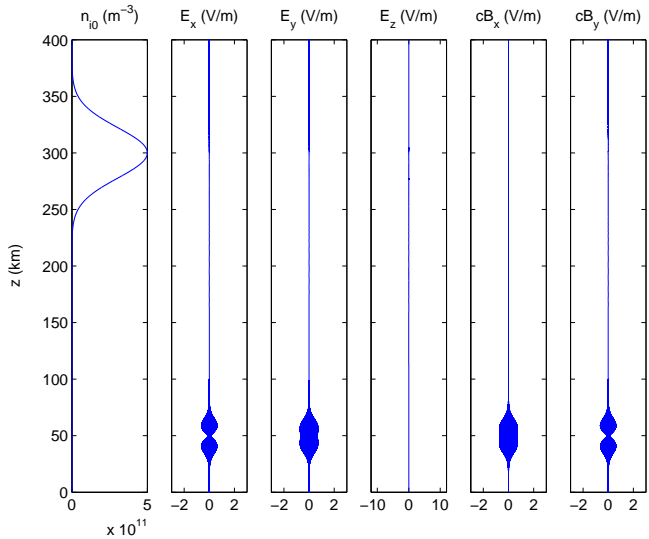}
\caption{The ion density profile, and the electric and magnetic
field components at time $t=1.752$ ms.} \label{Fig7}
\end{figure}

In the simulation, the EM pulse propagates without changing shape
through the neutral atmosphere, until it reaches the ionospheric
layer. At time $t=0.720$ ms, shown in Fig.~\ref{Fig2}, the EM pulse
has reached the lower part of the ionosphere. The initially linearly
polarized EM wave undergoes Faraday rotation due to the different
dispersion properties of the L-O and R-X modes (we have adopted the
notation ``L-O mode'' and ``R-X mode'' for the two high-frequency EM
modes, similarly as, e.g., \textit{Goertz and Strangeway} [1995]) in
the magnetized plasma, and the $E_y$ and $B_x$ components are
excited. At $t=0.886$ ms, shown in Fig.~\ref{Fig3-4}, the L-O and
R-X mode pulses are in the vicinity of their respective turning
points, the turning point of the L-O mode being at a higher altitude
than that of the R-X mode; see panel a) of Fig.~\ref{Fig3-4}. A
closeup of this region, displayed in panel b), shows that the first
maximum of the R-X mode is at $z\approx 270.5$ km, and the one of
the L-O mode is at $z\approx 277$ km. The maximum amplitude of the
R-X mode is $\approx 3$ V/m while that of the L-O mode is
$\approx10$ V/m; the latter amplitude maximum is in agreement with
those obtained by \textit{Thid\'e and Lundborg}, [1986], for a
similar set of parameters as used here. The electric field
components of the L-O mode, which at this stage are concentrated
into a pulse with a single maximum with a width of $\approx 200$ m,
are primarily directed along the geomagnetic field lines, and hence
only the $E_z$ and $E_x$ components are excited, while the magnetic
field components of the L-O mode are very small. At $t=0.948$ ms,
shown in panel a) of Fig.~\ref{Fig5-6}, both the R-X and L-O mode
wave packets have widened in space, and the EM wave has started
turning back towards lower altitudes. In the closeup of the EM wave
in panel b) of Fig.~\ref{Fig5-6}, one sees that the L-O mode
oscillations at $z\approx 277$ km are now radiating EM waves with
significant magnetic field components. Finally, shown in
Fig.~\ref{Fig7} at $t=1.752$, the EM pulse has returned to the
initial location at $z=50$ km. Due to the different reflection
heights of the L-O and R-X modes, the leading (lower altitude) part
of the pulse is primarily R-X mode polarized while its trailing
(higher altitude) part is L-O mode polarized. In the center of the
pulse, where we have a superposition of the R-X and L-O mode, the
wave is almost linearly polarized with the electric field along the
$y$ axis and the magnetic field along the $x$ axis. The direction of
the electric and magnetic fields here depends on the relative phase
between the R-X and L-O mode.

\begin{figure}[p]
\includegraphics[width=0.5\textwidth]{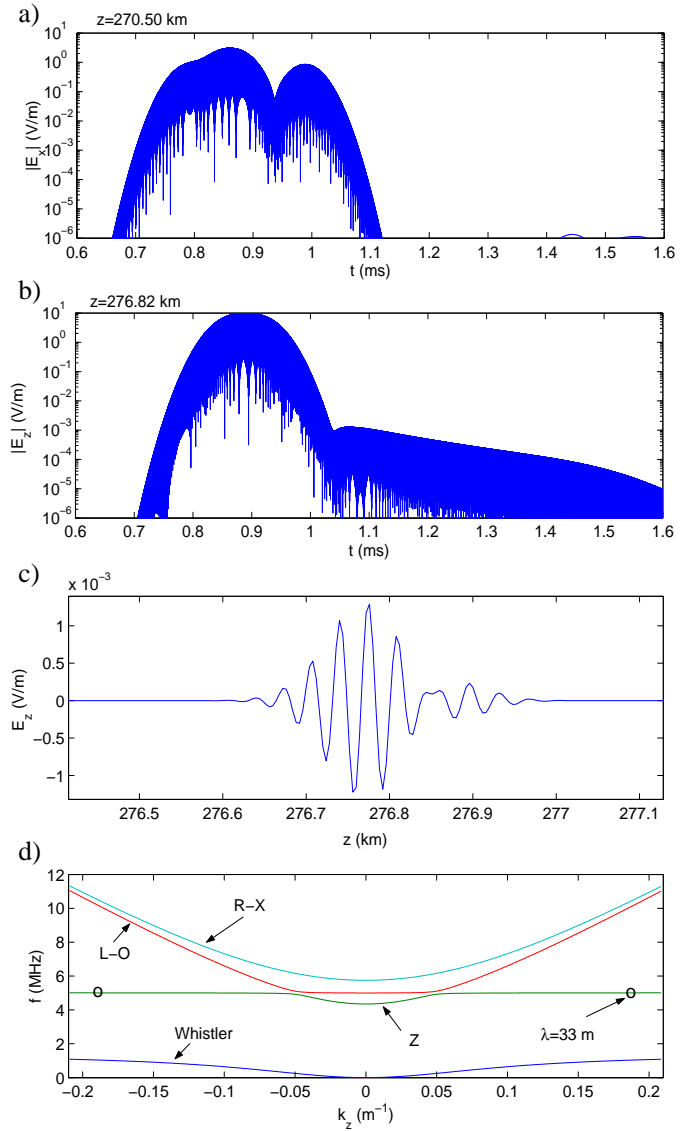}
\caption{ a) The amplitude of the electric field component $E_x$ at
$z=270.50$ km, near the turning point of the R-X mode, and b) the
amplitude of the electric field component $E_z$ at $z=276.82$ km,
near the turning point of the L-O mode. c) A snapshot of
low-amplitude electrostatic waves of wavelength $\lambda\approx 33$
m (wavenumber $k=2\pi/\lambda\approx 0.19\,{\rm m}^{-1}$), observed
at time $t=1.152$ ms, and d) Dispersion curves (lower panel)
obtained from the Appleton-Hartree dispersion relation with
parameters $\omega_{\rm pe}=31.4\times 10^6\,{\rm s}^{-1}$ (5 MHz),
$\omega_{\rm ce}=8.80\times 10^6\,{\rm s}^{-1}$ (1.4 MHz) and
$\theta=13^\circ=0.227$ rad. We identify the high-frequency R-X and
L-O modes, as well as the Z-mode which extends to the electrostatic
Langmuir/Upper hybrid branch for large wavenumbers; the circles
indicate the approximate locations on the dispersion curve for the
electrostatic oscillations shown in panel c). For completeness we
also show the low-frequency electron whistler branch in panel d).}
\label{Fig8-9}
\end{figure}

\begin{figure}[htb]
\includegraphics[width=0.48\textwidth]{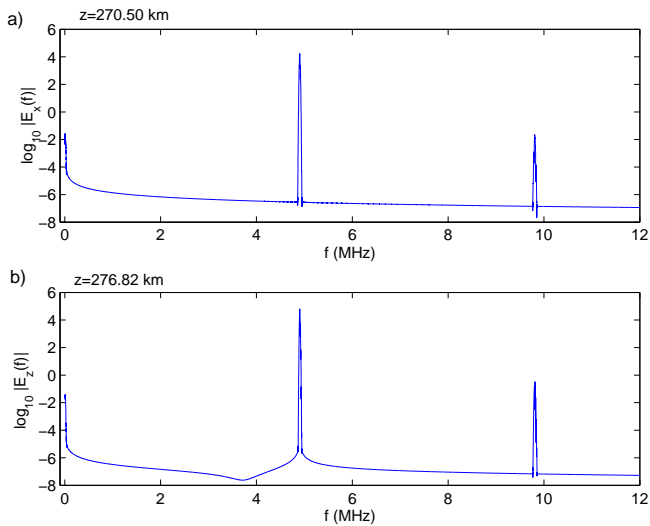}
\caption{The frequency spectrum (10-logarithmic scale) of a) the
electric field component $E_x$ at the altitude $z=270.50$ km, and b)
of $E_z$ at the altitude $z=276.82$ km.} \label{Fig10}
\end{figure}

In Fig.~\ref{Fig8-9}, panel a), we have plotted the electric field
component $E_x$ at $z=270.50$ km, near the turning point of the R-X
mode and in panel b) we have plotted the $E_z$ component at
$z=276.82$ km, near the turning point of the L-O mode. We see that
the maximum amplitude of $E_x$ reaches $3$ V/m at $t=0.87$ ms, and
that of $E_z$ reaches $10$ V/m at $t=0.9$ ms. The electric field
amplitude at $z=270.50$ km has two maxima, due to the L-O mode part
of the pulse, which is reflected at the higher altitude $z=276.82$
km and passes twice over the altitude $z=270.50$ km. We also observe
weakly damped oscillations of $E_z$ at $z=276.82$ km for times
$t>1.05$ ms, which decrease exponentially in time between $t=1.1$ ms
and $t=1.5$ ms as $E_z\propto\exp(-\gamma t)$ with
$\gamma=6.5\times10^3$ s$^{-1}$. We found from the numerical values
that $\gamma\approx c k_n/2$, where $k_n=d {\rm ln} n_{i0}/dz\approx
4.6\times 10^{-5}\,{\rm m}^{-1}$ is the inverse ion density scale
length at $z=277$ km, but we are not certain how general this result
is. No detectable magnetic field fluctuations are associated with
these weakly damped oscillations, and we interpret them as
electrostatic waves that have been produced by mode conversion of
the L-O mode. The amplitudes of the $E_x$ and $E_y$ components are
also much weaker than that of the $E_z$ component for these
oscillations. A closeup of these electrostatic oscillations at
$t=1.152$ ms is displayed in panel c) of Fig.~\ref{Fig8-9}, where we
see that they have a wavelength of approximately 33 m (wavenumber
$0.19\,{\rm m}^{-1}$). In panel d) of Fig.~\ref{Fig8-9}, we have
plotted the frequency $f=\omega/2\pi$ as a function of the
wavenumber $k$, where $\omega$ is obtained from the Appleton-Hartree
dispersion relation [\textit{Stix}, 1992]
\begin{equation}
\omega^2=c^2 k_z^2+\frac{2\omega_{\rm pe}^2(\omega^2-\omega_{\rm
pe}^2)}{2(\omega^2-\omega_{\rm pe}^2)-\omega_{\rm
ce}^2\sin^2\theta\pm \omega_{\rm
 ce}\Delta}.
 \label{Appleton}
\end{equation}
Here $\Delta=[\omega_{\rm
ce}^2\sin^4\theta+4\omega^{-2}(\omega^2-\omega_{\rm
pe}^2)^2\cos^2\theta]^{1/2}$, $\omega_{\rm pe}$ ($\omega_{\rm ce}$)
is the electron plasma (cyclotron) frequency, and $\theta$ is the
angle between the geomagnetic field and the wave vector ${\bf k}$,
which in our case is directed along the $z$-axis, ${\bf k}=k_z
\widehat{\bf z}$. We use $\omega_{\rm pe}=31.4\times 10^6\,{\rm
s}^{-1}$ (corresponding to $f_{\rm pe}=5$ MHz), $\omega_{\rm
ce}=8.80\times 10^6\,{\rm s}^{-1}$ (corresponding to $f_{\rm
ce}=1.4$ MHz) and $\theta=13^\circ=0.227$ rad. The location of the
electrostatic waves whose wavelength is approximately 33 m and
frequency 5 MHz is indicated with circles in the diagram; they are
on the same dispersion surface as the Langmuir waves and the upper
hybrid waves/slow Z mode waves with propagation parallel and
perpendicular to the geomagnetic field lines, respectively. The mode
conversion of the L-O mode into electrostatic oscillations are
relatively weak in our simulation of vertically incident EM waves,
and theory shows that the most efficient linear mode conversion of
the L-O mode occurs at two angles of incidence in the magnetic
meridian plane, given by, e.g., Eq. (17) in [\textit{Mj{\o}lhus},
1990].

The nonlinear effects at the turning point of the L-O and R-X modes
are investigated in Fig.~\ref{Fig10} which displays the frequency
spectrum of the electric field component $E_x$ at the altitude
$z=270.5$ km and of $E_z$ at the altitude $z=276.82$ km. The
spectrum shows the large-amplitude pump wave at 5 MHz and the
relatively weak second harmonics of the pump wave at 10 MHz at both
altitudes (the slight downshift is due to numerical errors produced
by the difference approximations used in space and time). Visible
are also low-frequency oscillations (zeroth harmonic) due to the
nonlinear down-shifting/mixing of the high-frequency wave field.

\section{Summary}

In conclusion, we have presented a full-scale numerical study of the
propagation of an EM wave and its linear and nonlinear interactions
with an ionospheric layer. We observe the reflection of the L-O and
R-X modes at different altitudes, the mode conversion of the L-O
mode into electrostatic Langmuir/upper hybrid waves as well as
nonlinear harmonic generation of the high-frequency waves. Second
harmonic generation have been observed in ionospheric heating
experiments [\textit{Derblom et al.}, 1989;
\textit{Blagoveshchenskaya et al.}, 1998] and may be partially
explained by the cold plasma model presented here.

{\bf Acknowledgment} This work was supported financially by the
Swedish Research Council (VR).

%\end{article}

\end{document}